\documentclass[reprint,aps,prl,twocolumn,amsfonts,amsmath,amssymb,superscriptaddress,notitlepage]{revtex4-2}
\usepackage[utf8]{inputenc}
\usepackage[T1]{fontenc}
\usepackage{lmodern}
\usepackage{hyperref}
\usepackage{xcolor}
\usepackage{graphicx}
\usepackage{titlesec}
\usepackage{placeins}
\usepackage{mathrsfs}

\definecolor{mygreen}{rgb}{0, 0.6, 0}

\usepackage{soul}
\setstcolor{blue}
\usepackage{dcolumn}
\usepackage{bm}
\begin{document}

\title{Force - Velocity Relationship in Branched Actin Networks: Consequences of Entanglement, Drag and Stall Force.}

\author{Magid Badaoui}
\author{Serge Dmitrieff}
\email{serge.dmitrieff@ijm.fr}
\affiliation{Université Paris Cité, CNRS, Institut Jacques Monod, F-75013, Paris}

\date{\today} 

\begin{abstract}
We investigate the growth of a branched actin network under load. Using a combination of simulations and theory, we show that the network adapts to the load and exhibits two regimes: a finite velocity at low stress, followed by a power-law decay of the velocity as a function of stress. This decay is explained by a theoretical model relating branched network elasticity to filament entanglement. The finite maximum velocity is attributed to network drag, which dictates dynamics at low stress. Additionally, analysis of filament stall force contribution reveals a transition from a stalled network to a growing network, when the filament stall force exceeds a critical value controlled by the applied stress.
\end{abstract}

\maketitle

\textit{Introduction.}---Actin, a central component of the cytoskeleton and essential for cell motility, generates forces through motor-induced contractility, polymerization, or depolymerization \cite{Blanchoin2014}. Polymerization drives essential processes such as endocytosis, lamellipodium protrusion, and intracellular pathogen propulsion \cite{Weinberg2012Clathrin-mediatedYeast,Small1995ActinLamellipodium,Cossart1994TheMonocytogenes}. Powered by the chemical potential difference between free monomers and those bound to filaments, this process converts chemical energy into mechanical force \cite{Kovar2004InsertionalForces}. The Brownian ratchet model explains how, in the presence of a membrane or wall, thermal fluctuations enable monomer addition to the barbed end of a filament. However, as the opposing wall force on the filament increases, growth exponentially slows and eventually stalls at a few piconewtons \cite{Hill1981, Peskin1993, Footer2007DirectTrap}. Extending this single-filament model to dense actin networks is challenging due to their complex mechanics and architecture.

Branched actin networks, with their dendritic architecture, exhibit a high degree of mechanical adaptability under stress. Advances in understanding the nucleating Arp2/3 complex and its activation by Nucleation Promoting Factors (NPFs) have enabled the production of \textit{in vitro} structures that closely resemble dense, branched networks found \textit{in vivo} \cite{Rohatgi1999TheAssembly, Loisel1999ReconstitutionProteins}. These networks stiffen as stress increases due to structural remodeling, with their elastic modulus growing proportionally to the stress \cite{Marcy2004, Chaudhuri2007, Bieling2016ForceNetworks, Bauer2017ANetworks, Li2022}. Bieling et al. \cite{Bieling2016ForceNetworks} studied the steady-state growth of actin columns opposing an Action Force Microscopy (AFM) tip. They observed that higher growth stress increases network density while velocity decreases with a convex curve that deviates from the predictions of the Brownian ratchet model. Recent findings emphasize the complex interplay of molecular processes that govern network growth and stability, such as filament elongation, capping and nucleation \cite{Funk2021,Li2022,Lappalainen2022}.

Despite these advances, the stress adaptation of branched networks remains poorly understood. Here we present a theory predicting the force-velocity relationship of growing actin network, taking into account the role of filament stall force, that we can verify with simulations. 

\begin{figure}[t]
\centering
\includegraphics[width=\columnwidth]{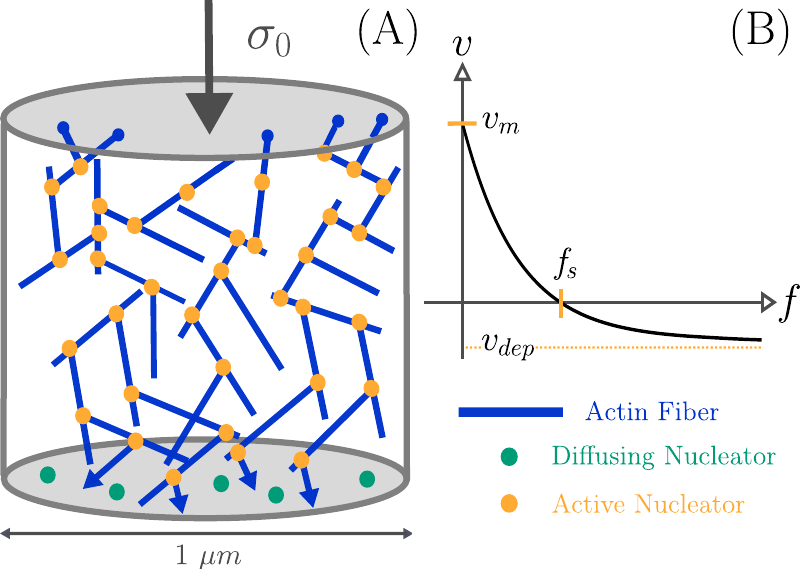}
\caption{(A) Schematic of the simulation of an actin network growing under stress $\sigma_0$. Network expansion is maintained by a constant renewal rate of diffusing nucleators: $k$. These nucleators generate new branches when they encounter a filament. A simulation movie can be found in supplemental material. (B) Force-velocity curve for a single filament based on the Brownian ratchet model.}
\label{fig:fig_1}
\end{figure}

We find that actin networks grow at a load-dependent velocity. At high loads, the velocity is a power-law of stress, that can be explained through simple scaling laws. Moreover, there exists a critical stall force - scaling linearly with the applied stress -below which the network cannot grow. At low loads, the velocity is limited by the network drag, and exhibits a power-law scaling with time.
\medbreak
\textit{Stress adaptation of networks at high loads.}---To understand the stress adaptation of networks under load, we simulate actin growing in a piston where an external stress $\sigma_0$ is applied on the lid, at the piston top. Active nucleators are regularly added at a surface rate $k$, on the bottom disk, and diffuse on it. When a diffusing nucleator encounters a polymerized filament, it binds to it and initiates the growth of a new polymerizing filament with a branching angle of 70° between the mother and the daughter branches, as shown in FIG. \ref{fig:fig_1}(A). This setup mimics branched actin nucleation by Arp2/3, and its activation by WASP family proteins on cellular membranes. This configuration is close to existing experimental setups \cite{Bieling2016ForceNetworks, Li2022}. Here we assume filaments to grow up to a maximum length $\ell_m$, mimicking the effect of capping proteins. Indeed, it was shown that the average filament length remains independent of the applied stress, since capping is force dependent in the same way as actin, following a Brownian ratchet mechanism \cite{Li2022}.

While branched networks would be predicted to exhibit vanishing elasticity due to their low connectivity \cite{Maxwell1864OnFrames}, recent work propose that contact points between filaments effectively increase the coordination number, and thus the elastic modulus \cite{Bouzid2024TransientActin}. In this context, the distance between contact points $\xi$, scales like:
\begin{eqnarray}\label{martin_xi}
    \xi \approx {\left( \frac{\kappa}{\sigma_0} \right)}^{2/7}r^{-1/7}
\end{eqnarray}
Where $\kappa$ is the bending stiffness of the filament, and $r$ its radius. This entanglement theory holds primarily in dense regimes where $\xi$ is much smaller than the filament’s maximum length, $\ell_m$, making $\xi$ the characteristic length scale in the system. The authors then find a scaling of density as a function of stress \cite{Bouzid2024TransientActin}:
\begin{eqnarray}\label{martin_rho}
\rho_s \propto \kappa^{-2/7}\sigma_{0}^{2/7}r^{8/7}
\end{eqnarray}
At timescales larger than filament growth, the total flux of actin $j$ is stationary and independent of the stress (see Supp FIG. S8), and reaches $j = k \ell_m s_0$, $s_0$ being the cross section of a filament. Because of mass conservation, the stationary growth velocity should be $v_s \propto j / \rho_s$, allowing us to derive the growth velocity scaling:
\begin{eqnarray}\label{martin_v}
    v_s \propto j\kappa^{2/7}\sigma_{0}^{-2/7}r^{-8/7}
\end{eqnarray}
To test this result, we performed simulations of the system with Cytosim \cite{Nedelec2007}. Implementation details are in Supp Sec I.A . In short, filament are modeled as lines discretized by regularly-spaced vertices. Each vertex follows an over-damped Langevin equation, that is solved using a semi-implicit scheme. Besides thermal fluctuations, forces on the vertices stem from filament bending elasticity, steric repulsion between filaments, and confinement by the piston. Filament bending energy follows the standard elasticity of thin rods: $e_b \propto \kappa C^2 /2$ (with $\kappa$ the bending modulus and $C$ the curvature). Steric repulsion between filaments is assumed to follow Hooke's law: $e_s \propto (d - 2 r)^2$, with $d$ the distance between the filaments, when the filaments are in contact ($d<2r$). The confinement potential is also quadratic, with $e_c \propto (z-z^*)^2$ when filaments are outside the piston (with $z$ the vertex position along the vertical axis and $z^*$ the closest boundary). 

We first consider filaments growing at a constant speed $v_m$ until their maximal length $\ell_m$. This is equivalent to considering a limit case of the Brownian ratchet model, where the stall force $f_s$ is infinite, see FIG. \ref{fig:fig_1}(B). Active nucleators binding filaments instantly generate a new branch with a favorite angle 70° and a high angular stiffness. New nucleators are generated on the bottom disk with a surface rate $k$. The piston lid is moved with time to apply a constant stress $\sigma_0$, as detailed in Supp Sec I.A.1. The network is primed with five initial filaments of length $\ell_m$ to initiate the nucleation reaction. Main simulation parameters are in TABLE I, simulation details can be found in Supp Sec I.B along with a configuration file in supplemental material.
\medbreak
\begin{table}[h!]
    \centering
    \begin{tabular}{|c|c|c|}
        \hline
        $\ell_m$ = 0.3 $\mu$m & $r$ = 0.005 $\mu$m & $\kappa$ = 0.004 pN.$\mu$m\textsuperscript{2}  \\ \hline
        $v_0$ = 1 $\mu$m.s\textsuperscript{-1} & $v_{dep}$ = 0.1 $\mu$m.s\textsuperscript{-1} & $k S$ = 500 s\textsuperscript{-1}  \\ \hline
    \end{tabular}
    \caption{Main simulation parameters values.}
\end{table}

\begin{figure*}[t]
\centering
\includegraphics[width=\textwidth]{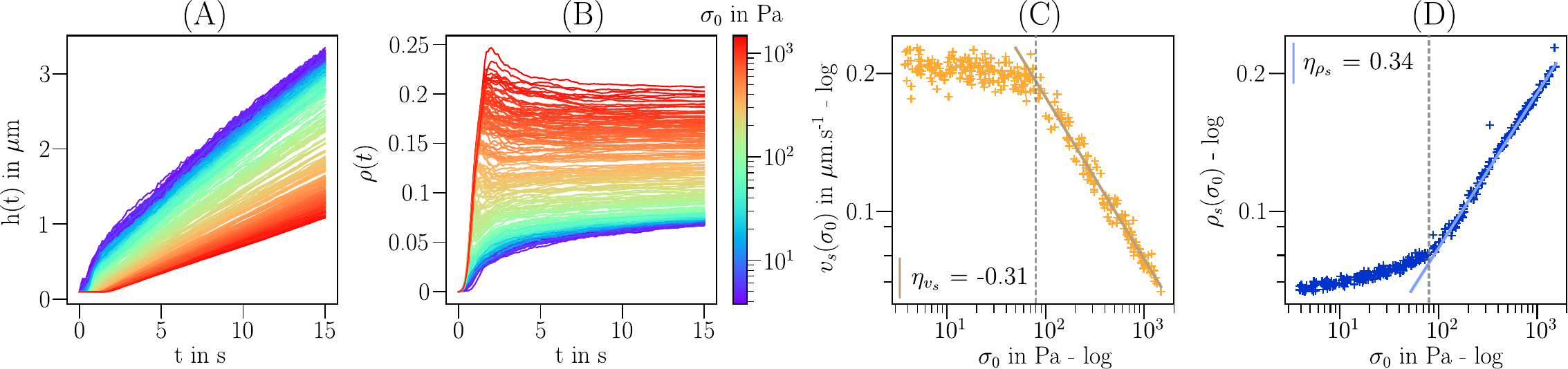} \caption{(A) Network height $h(t)$ and (B) density $\rho(t)$ as a function of time $t$ for various stresses $\sigma_0$ with $f_{s} = \infty$ pN. (C) Stationary velocity $v_s$ and (D) density $\rho_s$ as a function of stress, both shown in log-log scales. A power-law behavior is observed for high stress, following a near-constant phase for both observables. The gray dotted-line indicates the transition between the two phases, here at 80 Pa. 300 simulation points.}
\label{fig:fig_2}
\end{figure*}

\textit{Numerical results.}---We performed multiple simulations, varying the applied stress $\sigma_0$ while assuming the filament stall force to be infinite. At high stress, following a transition period, the system reached a growing steady-state as shown in FIG. \ref{fig:fig_2}(A), with no nucleators accumulation (see Supp Sec I.C). Notably, the network's growth velocity decreases as the applied stress increases. Simultaneously, network density also reaches a stationary state. The density plateau value increases with the applied stress, as shown in FIG. \ref{fig:fig_2}(B). However, at low applied stress, the density does not saturate with time. Because of the conservation of actin flux, this means that the velocity does not quite saturate.
Accordingly, the dependence of the velocity on the applied stress shows two distinct phases: for stresses below approximately 80 Pa, the velocity is nearly independent of the applied stress. In the high-stress regime, the velocity exhibits a power-law decay, scaling as $v_s \propto \sigma_{0}^{\eta_{v_s}}$ where $\eta_{v_s}$ = -0.31, FIG. \ref{fig:fig_2}(C). Hence, at high stress, density scales with stress as $\rho_s \propto \sigma_{0}^{\eta_{\rho_s}},$ where $\eta_{\rho_s}$ = 0.34, ensuring that the flux ($j = \rho_s v_s$) remains near-independent of the applied stress, FIG. \ref{fig:fig_2}(D). This is in agreement with our theoretical predictions $\rho_s \propto \sigma_{0}^{2/7}$ and $v_s \propto \sigma_{0}^{-2/7}$, with $2/7\approx0.29$. Indeed, at high stress $\sigma_0$ > 80 Pa, we have $\xi$ < 0.13 $\mu$m, smaller than filament size $\ell_m$ = 0.3 $\mu$m, and the entanglement theory should hold. Details about $v_s$ and $\rho_s$ estimations are in Supp Sec II.A.
\medbreak
\textit{Low stress regime.}---At low applied stress, network growth speed is still much smaller than the filament growth speed $v_m$ = 0.9 $\mu$m.s\textsuperscript{-1}. We hypothesized that the drag of the network could become dominant in opposing network growth and limiting the growth speed. In Cytosim, where complex hydrodynamic interactions are absent, network drag is merely the sum of individual filament drags. Following the model from \cite{Tirado1979TranslationalCylinders}, considering an angular average, this cumulative drag can be expressed as:
\begin{eqnarray} \label{gamma}
    \Gamma(t) = \sum_{i=1}^{N(t)} \frac{3 \pi \eta \ell_{i}(t)}{\ln \big(\frac{\ell_{i}(t)}{2r}\big) + 0.312}
\end{eqnarray}
Where $\eta$ represents the medium dynamic viscosity, $\ell_i$(t) the length of filament $i$ at time $t$, $r$ the filament radius and $N(t)$ the number of filaments at time $t$. For low applied stress $\sigma_0 \le$ 80 Pa, we observed that the drag-induced stress $\sigma_{d}(t)$ = $\Gamma(t) v_{s}(\sigma_0)$ quickly grew larger than $\sigma_0$, as shown in Supp Sec III.A. Therefore, as the network expands, newly formed slices encounter an increasing amount of actin, leading to a rise of density of these slices. Furthermore, over extended
time intervals, slices are expected to undergo elastic relaxation, which can be inferred from a density decrease as time progresses, as seen in Supp Sec III.B. Thus, the total growth speed of the network $v_\tau$ should come from both polymerization and network expansion:
\begin{eqnarray}\label{free_growth_model}
    v_{\tau}(t) = v(t) + h(t) \bar{\dot{\epsilon}}(t) \ ; \ \bar{\dot{\epsilon}}(t) = \int_{0}^{h(t)}\frac{dx}{h(t)}\partial_{t}\epsilon(x,t)
\end{eqnarray}
Here, $v(t)$ is the polymerization-driven velocity, and $ \bar{\dot{\epsilon}}(t)$ is the average elastic relaxation rate across the network. Although density relaxation occurs, it can reasonably be neglected over the observed time scales (see Supp Sec III.B) and we can use the approximation $v_{\tau}(t) \simeq v(t)$.

\begin{figure}[h]
\centering
\includegraphics[width=\columnwidth]{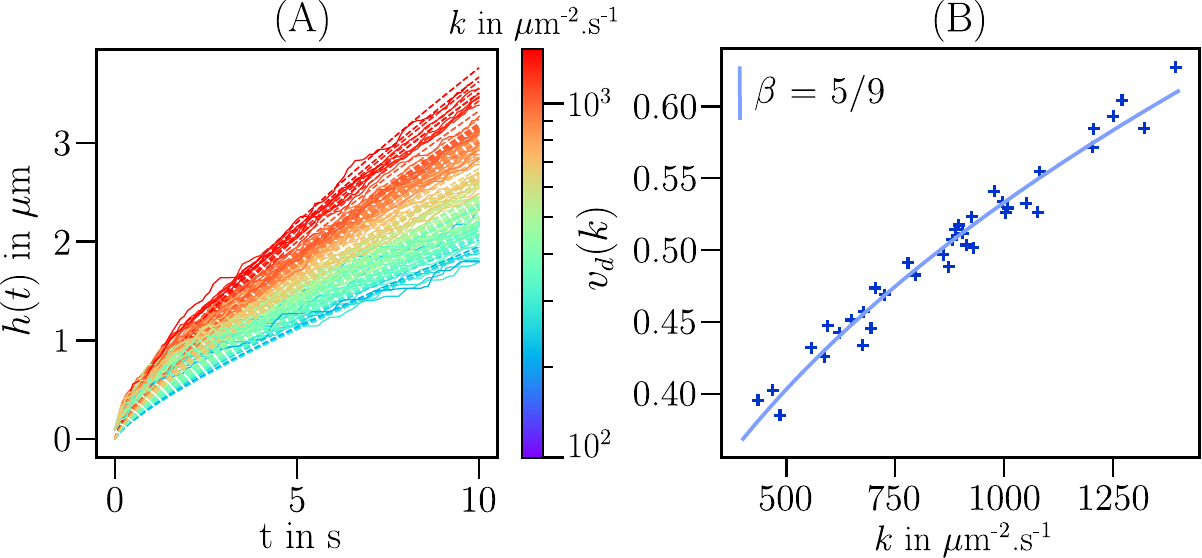}
\caption{(A) Height h(t) over time with power-law fit (dotted line) for various nucleator rate $k$ and $f_{s}$ = $\infty$ pN. (B) Pre-factors of the power-law $v_d$ as a function of $k$ with exponent 5/9 as predicted by theory.}
\label{fig:fig_3}
\end{figure}

In this framework, the system is modeled as polymerizing filaments that push against a gradually increasing drag-induced stress, denoted by $\sigma_d$(t). As filaments reach their maximum length, the drag-velocity relationship simplifies to: $\sigma_d(t) = \gamma_0 N(t)v(t)/S$, where $\gamma_0$ is the drag for a filament of length $\ell_m$, $S$ is the cross section area of the cylinder and $N(t)$ is the number of actin filaments ($N(t)$=$kSt$). Assuming we are within the entanglement regime, we use equation (\ref{martin_v}): $v(t) = j\kappa^{2/7}r^{-8/7}\sigma_d(t)^{-2/7}$. From these relationships, we derive the scalings of $v(t)$ and $\rho(t)$ in terms of $k$ (detailed calculation in Supp Sec III.C):
\begin{eqnarray}\label{free_growth_scaling}
    v(t) &\propto& v_{d}(k)t^{-2/9} \quad ; \quad v_{d }(k) \propto k^{5/9} \\
    \rho(t) &\propto& \rho_{d}(k)t^{2/9} \quad \ \; ; \quad \rho_{d}(k) \propto k^{4/9}
\end{eqnarray}
This implies:
\begin{eqnarray}\label{free_h_scaling}
    h(t) \propto \frac{9}{7} v_{d}(k) t^{7/9}
\end{eqnarray}
To verify these predictions, we simulated the system without external stress $\sigma_0 = 0$, varying systematically $k$. In FIG. \ref{fig:fig_3}(A), we present power-law fits for $h(t)$. When the nucleation rate is high enough, $k\ge$ 200 $\mu\text{m}^{-2}.\text{s}^{-1}$, we find a good agreement between the predicted power-law and simulations. We can also fit the corresponding network growth velocity $v(t) \propto v_{d}(k) t^{-2/9}$ ; the prefactor $v_{d}(k)$ indeed follows the scaling law $v_{d}(k) \propto k^{5/9}$ predicted by our theory, see FIG. \ref{fig:fig_3}(B) (corresponding density $\rho(t)$ plot in Supp Sec III.D). Therefore, at zero applied stress, entanglement theory explains the time evolution of growth velocity and network density, provided that the nucleation rate $k$ is high enough to allow for a high drag-induced stress.

At low $k$ values, deviations from entanglement theory and notable elastic relaxation effects can occur. Indeed, networks with lower $k$ require more time to develop sufficient drag for enhanced entanglement, and exhibit more irregular structures and greater density relaxation over time, as shown in Supp FIG. S11.

\medbreak

\textit{Stall force dependence.}---Actual biological polymers such as actin have a maximum load against which they can grow, called \emph{stall force} $f_s$. This is determined by the ratio between the polymerization Gibbs free energy $\delta G$ and the monomer size $a$: $f_s = - \delta G /a$. A consequence of this is that filament growth speed should depend exponentially on the force, FIG. \ref{fig:fig_1}(B) \cite{Hill1981,badaoui2023multi}:
\begin{gather}\label{brownian_racthet}
    v = v_0e^{-\frac{f}{f_0}} - v_{dep} \quad ; \quad f \ge 0 \\
    f_s = f_0 \log{\left( \frac{v_0}{v_{dep}} \right)}
\end{gather}
With $f_0$ the growing force, $v_0$ the polymerization velocity and $v_{dep}$ the de-polymerization velocity. Here we make the common assumption that $v_{dep}$ is independent of $f$. We wondered wether our theory could also predict the mechanics and growth velocity of networks with finite stall force. We ran similar simulations to FIG. \ref{fig:fig_2} using a realistic finite stall force $f_{s} \approx$ 3.87 pN; see Supp Sec IV.A.

\begin{figure}[t]
\centering
\includegraphics[width=\columnwidth]{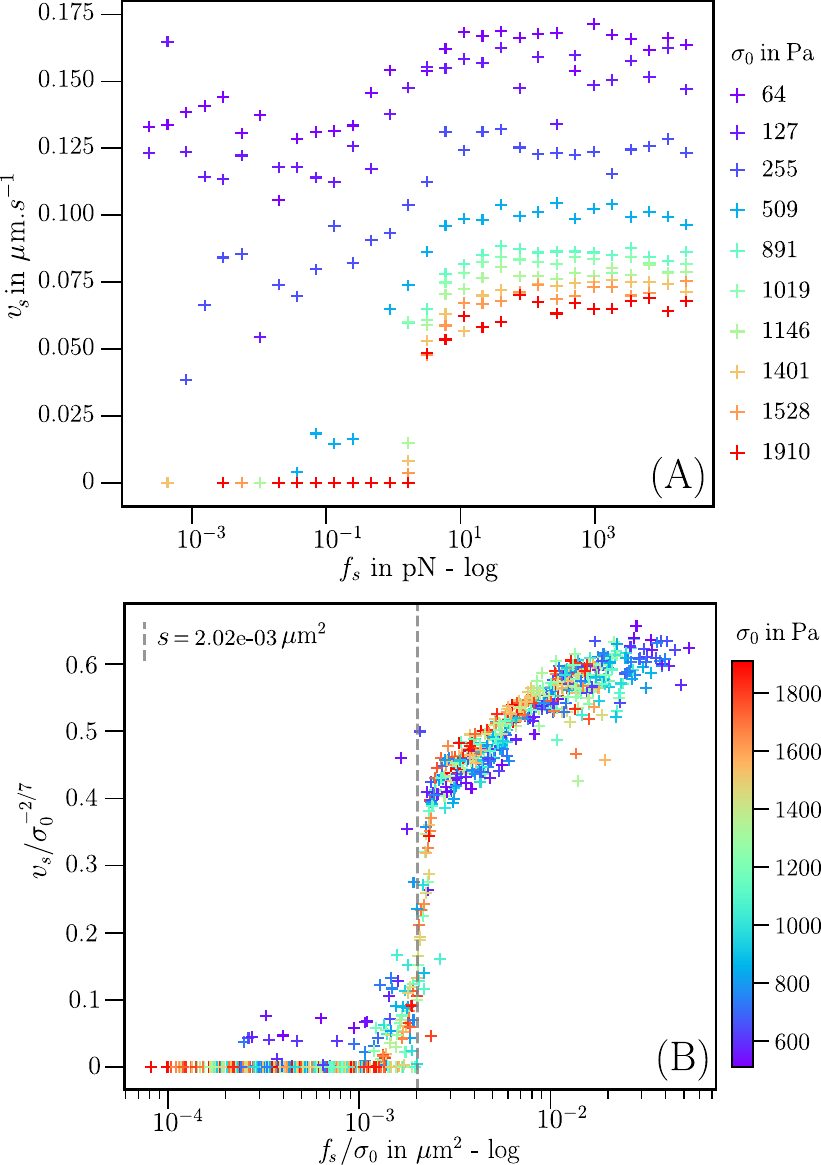}
\caption{(A) Stationary velocity $v_s$ as a function of stall force $f_s$ in x-log. Appearance of a stall force transition threshold for different stress values. 300 simulation points. (B) Re-scaled stationary velocity $v_s$ with respect to entanglement theory versus stall force over stress for different stress values. The gray dotted-line represents the effective cross section transition $s$. 1000 simulation points.}
\label{fig:fig_4}
\end{figure}

These simulations displayed similar behavior as before: a constant growth phase followed by power-law decay (see Supp FIG. S13), but with a different power-law exponent. We therefore needed to include the finite stall force in the theory. The energy gain by polymerization is the polymerization free energy $\delta G = - f_s a$ while the cost is $\sigma_0 \delta V$, with $\delta V$ the effective volume per monomer. Therefore, polymerization will happen spontaneously only if $f_s \sigma_0 > s$, with $s$ the effective cross-section of a filament. When polymerization occurs, we do expect entanglement theory to hold and the velocity should scale like $\sigma_{0}^{-2/7}$. Therefore, in the simplest limit, we expect (with $\Theta$ the Heaviside step function):
\begin{align}\label{poly_transi}
v_s \propto \sigma_{0}^{-2/7} \Theta \left( \frac{f_s}{\sigma_0} - s \right)
\end{align} 
An extensive series of simulations across a broad range of stall force and stress values confirmed the presence of an activation threshold, in FIG. \ref{fig:fig_4}(A). In particular, once the threshold is exceeded, velocity seems to become independent of stall force. To refine our understanding, we conducted a second series of simulations within a narrower parameter window, FIG. \ref{fig:fig_4}(B). We found that all simulation results remarkably aligned on a master curve, with growth speed behaving as a step function of $f_s / \sigma_0$ as predicted in Eq. \ref{poly_transi}. This step function is not quite a Heaviside function, hinting at higher order, more subtle interactions between filament stall force and network growth speed. Moreover, our theory could not explain the effective cross section $s \approx 2.0 \times 10^{-3} \mu\text{m}^2$ revealed by our simulations. Nonetheless, when plotted as a function of $f_s / \sigma_0$, the growth speed $v_s$ did scale with $\sigma_0^{-2/7}$ as expected, showing that entanglement theory still holds for filaments with finite stall force.

\medbreak

\textit{Discussion.}---In this study, we showed that we can extend entanglement theory to predict the force-velocity relationship of branched networks growing under pressure. When stall force is infinite, we highlight two regimes. At high applied stress, growth speed and network density reach a steady state and behave as power-laws of the applied stress. At low applied stress, drag stress dominates, and thus density and growth speed depend on time following a power-law. Moreover, the velocity scales as a power law of the nucleation rate. When the stall force is finite, there is a threshold of stall force above which network growth can occur - but the power-law behaviors of density and velocity at high stress are still valid. However, what sets the effective cross-section of actin filaments at the transition remains to be identified.

From a physical perspective, our simplified treatment of drag could be improved by incorporating hydrodynamic interactions, as recent models have done by treating filament ensembles as porous media \cite{Nazockdast2017CytoplasmicPositioning, Nazockdast2017CytoplasmicPositioningb}. While such refinements could enhance biological realism, they are unlikely to alter the fundamental mechanisms of stress adaptation observed here. Additionally, the entanglement framework does not fully account for polymerization or branching when the contact distance $\xi$ is smaller than the filament length. The spacing between branching points - potentially analogous to a characteristic length scale $\xi_{Arp2/3}$, may become a critical parameter, decreasing as filament density increases.

From a biochemical perspective, our simulation involves significant simplifications. Notably, the capping rate decreases under applied stress \cite{Li2022}, which leads to an increase in the number of free (uncapped) filament ends. These free ends have been experimentally shown to inhibit Arp2/3 actin nucleation \cite{Funk2021}, suggesting that nucleation rates should decrease as stress increases. Incorporating this effect in future simulations could enable a more direct comparison with experimental observations.

In two dimensions, a phase transition has been observed, where filament orientation strongly depends on the network growth velocity \cite{Weichsel2010}. While our simulations indicated an apparently random orientation, we suggest that further investigations are necessary to confirm the absence of a phase transition. The high prevalence of entanglement in such networks may act as a constraint, limiting the emergence of orientational order.

\medbreak

\begin{acknowledgments}
The authors would like to thank Olivia du Roure, Julien Heuvingh and Martin Lenz for fruitful scientific discussion. We also would like to thank Joel Marchand, Olivier Kirsh and Alix Silvert for the IT and cluster management support. Lastly, we thank Nicolas Minc and the entire team for the discussions and the scientific context. S.D. acknowledges financial support from INSERM Aviesan for the "MMINOS" project.
\end{acknowledgments}

\end{document}


\title{Supplemental Material: Force - Velocity Relationship in Branched Actin Networks: Consequences of Entanglement, Drag and Stall Force.}

\author{Magid Badaoui}
\author{Serge Dmitrieff}
\email{serge.dmitrieff@ijm.fr}
\affiliation{Université Paris Cité, CNRS, Institut Jacques Monod, F-75013, Paris}
\date{\today}

\maketitle
\tableofcontents
\newpage

\section{Simulation model}

\subsection{Numerical AFM, confinement and steric stiffness}

\subsubsection{Numerical AFM description}

Here we describe how we apply a near-constant stress $\sigma_0$ using a numerical equivalent of an atomic force microscope (AFM). Filaments are confined into a piston by a quadratic potential of stiffness $K_\text{confine}$, applied only on vertices outside the piston. Focusing on the piston axis ($z$-axis), each vertex $i$ feels a confinement force:
\begin{eqnarray}
f_i &=& K_\text{confine} (z_\text{bottom} - z_i) \quad \text{if} \quad z_i < z_\text{bottom} \\
f_i &=& K_\text{confine} (z_\text{top} - z_i) \ \quad \quad \text{if} \quad z_i > z_\text{top}
\end{eqnarray}
At each time step, we measure the stress applied on filaments by the lid of the piston $ \sigma_{top}$, as the sum of confinement forces by the lid divided by the lid surface area $S$:
\begin{eqnarray}
\sigma_\text{top} = - \sum^i_{z_i > z_\text{top}} \frac{f_i}{S}
\end{eqnarray}

Also, at each time step, we compare this stress to the target stress $\sigma_0$ by computing $\Delta \sigma = \sigma_{top} - \sigma_0$. Because $\partial_z \sigma_\text{top} \approx K_\text{confine} / S$, we change the lid position by $\delta z$ to converge towards $\sigma_0$, with: 
\begin{eqnarray}
 \delta z = \frac{\Delta \sigma S}{\alpha K_\text{confine}} 
\end{eqnarray}
With  $\alpha$ an attenuation parameter, set to $1$ unless mentioned otherwise (see FIG \ref{fig:s_4}). We limit $\| \delta z \|$ to a maximal value $\delta z^{max}$ to prevent instabilities.
We start the piston lid at a position $l_{eq} = 0.1 \ \mu \text{m}$, which corresponds to the height of the cylinder at $t$ = 0 s. We prevent the piston from shrinking below $l_{eq}$ to avoid collapse at early times.

\subsubsection{Confinement potential stiffness}
\begin{figure}[h]
    \centering
    \includegraphics[width=0.8\linewidth]{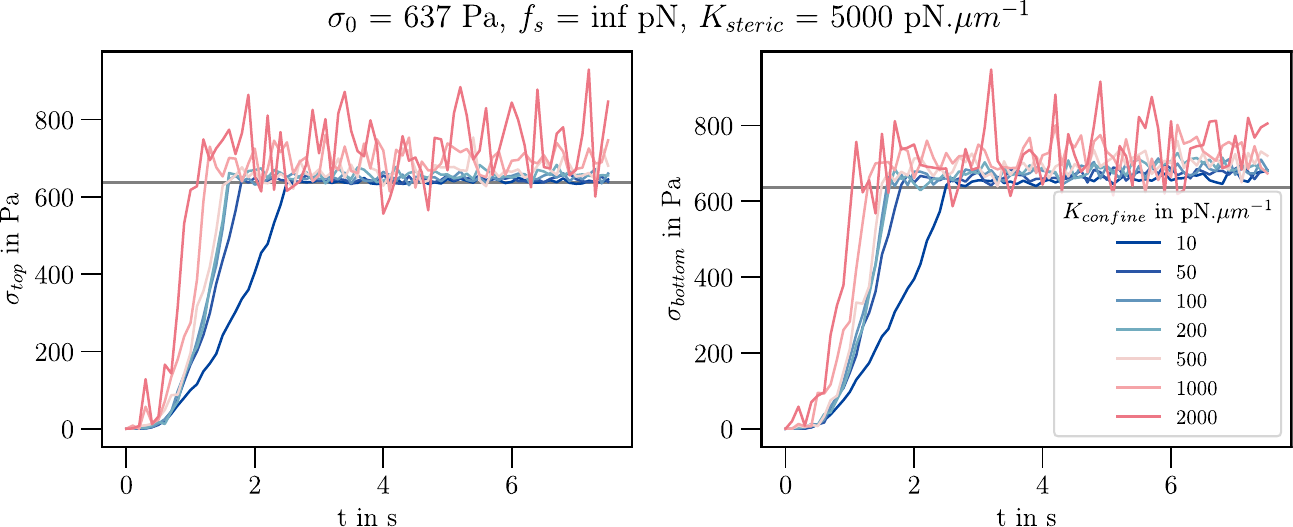}
    \caption{\centering{Stress measured by penetration length computation for top $\sigma_{top}$ (Left) and bottom $\sigma_{bottom}$ (Right) as a function of time $t$ for various $K_{confine}$, with: $\sigma_0$ = 637 Pa, $f_{s} = \infty$ pN and $K_{steric}$ = 5000 pN.$\mu$m\textsuperscript{-1}. The horizontal grey line is the target stress $\sigma_0$.}}
    \label{fig:s_1}
\end{figure}

We measured the confinement stress $ \sigma_{top}$ at the top $ \sigma_{top}$ and at the bottom $ \sigma_{bottom}$ as a function of time for various  values of the confinement stiffness $K_{confine}$, see FIG. \ref{fig:s_1}. When $K_{confine}$ is too low the network takes more time to reach its steady-state as more filaments are required to apply a stress approaching $\sigma_0$. When $K_{confine}$ is too high, the system reaches the steady-state quicker but looses in precision and fluctuates more. In the article, we used $K_{confine}$ = 1000 pN.$\mu$m\textsuperscript{-1} unless mentioned otherwise.

\FloatBarrier

\begin{figure}[h]
    \centering
    \includegraphics[width=\linewidth]{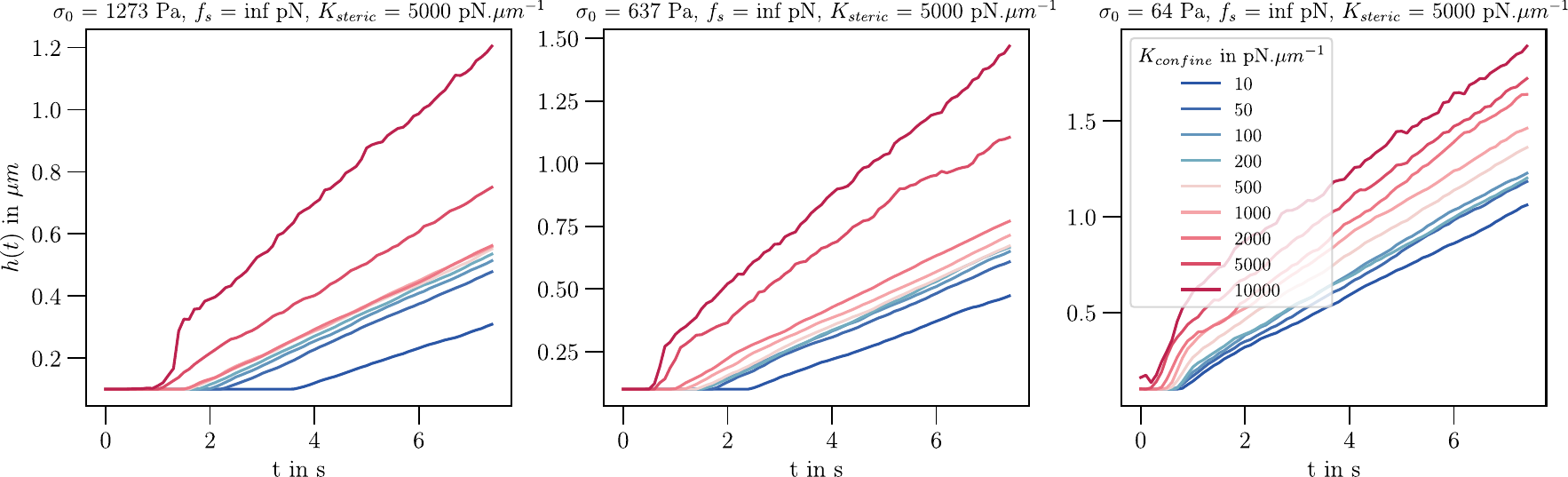}
    \caption{\centering{Height $h(t)$ as a function of time $t$ for various confinement stiffness parameters $K_{confine}$ under three stress $\sigma_0$ = 1273 Pa (Right), 636 Pa (Center) and 64 (Right) with $f_s = \infty$ pN, $K_{steric} = 5000 \ \text{pN}.\mu\text{m}\textsuperscript{-1}$.}} 
    \label{fig:s_2}
\end{figure}

We then measured the piston height $h(t)$ as a function of time for various values of $K_{confine}$, in FIG. \ref{fig:s_2}. We find that when the applied stress is high enough, $h(t)$ does not depend much on the confinement stiffness provided it is in the range $50 - 2000 \ \text{pN}.\mu\text{m}\textsuperscript{-1}$.

\FloatBarrier

\subsubsection{Steric potential stiffness}
\begin{figure}[h]
    \centering    \includegraphics[width=0.65\linewidth]{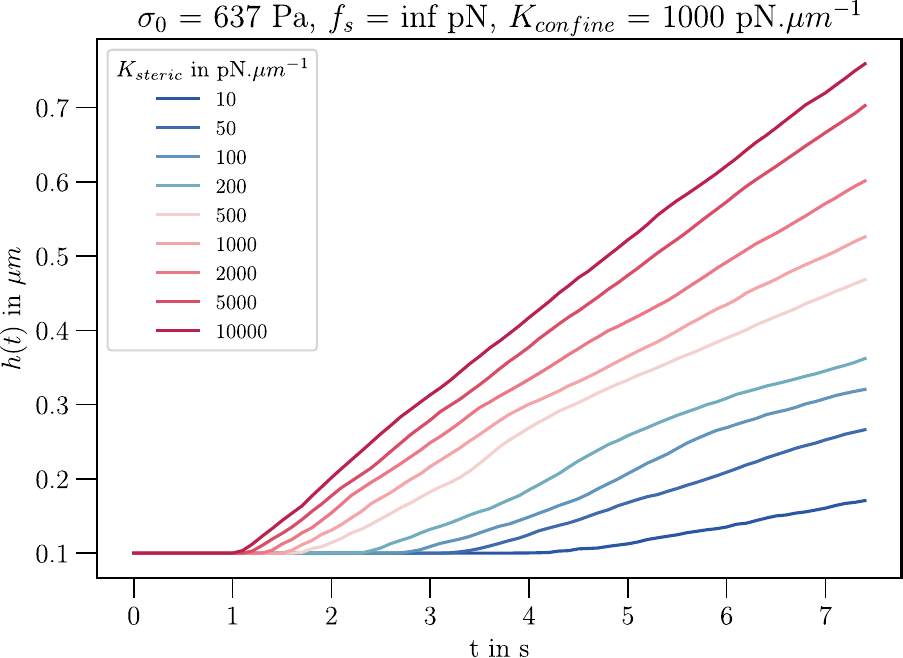}
    \caption{\centering{Height $h(t)$ as a function of time $t$ for various steric stiffness $K_{steric}$, with: $\sigma_0$ = 637 Pa, $f_s = \infty$ pN and $K_{confine}$ = 1000 pN.$\mu$m\textsuperscript{-1}.}}
    \label{fig:s_3}
\end{figure}

The steric repulsion between filaments is a soft, elastic potential. We expect the value of the steric stiffness $K_{steric}$ to influence on our results, with a limit $K_{steric} \rightarrow \infty$ corresponding to hardcore repulsions. We measured the height of the piston lid as a function of time for various values of $K_{steric}$ (see FIG. \ref{fig:s_3}), and we found that  $K_{steric} = 5000 \ \text{pN}.\mu \text{m}\textsuperscript{-1}$ was a good compromise between a very stiff repulsion and numerical performance.

\FloatBarrier

\newpage
\subsubsection{Attenuation parameter}
\begin{figure}[h]
    \centering
    \includegraphics[width=0.6\linewidth]{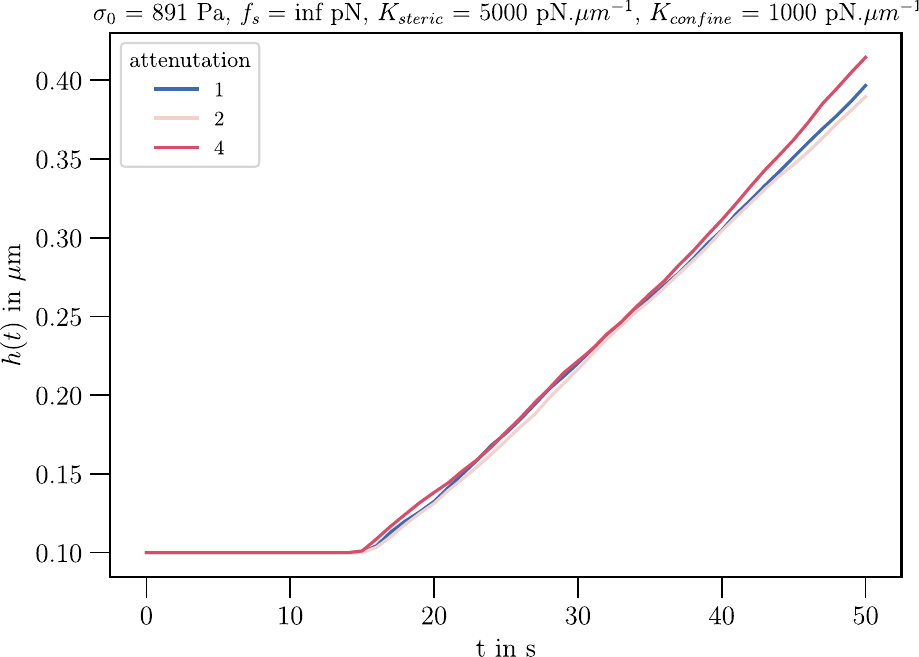}
    \caption{\centering{Height $h(t)$ as a function of time $t$ for various attenuation parameters $\alpha$ with: $\sigma_0$ = 891 Pa, $f_{s} = \infty$ pN, $K_{steric}$ = 5000 pN.$\mu$m\textsuperscript{-1} and $K_{confine}$ = 1000 pN.$\mu$m\textsuperscript{-1}.}} 
    \label{fig:s_4}
\end{figure}

We tried several values of the attenuation factor $\alpha$ and found that it does not impact the behavior of our simulation, see FIG. \ref{fig:s_4}.

\FloatBarrier

\subsection{Advanced parameters}

Here we will review the main elements of the configuration file that setup the parameters of a simulation. $k_{B}T = 0.0042$ pN.$\mu$m accounts for $T \simeq 304$ K. The viscosity $\eta$ is set to 0.1 Pa.s, which is in accordance with most experimental measures: the viscosity of the cytoplasm being a few hundred times the viscosity of water ($\eta_{water} \approx$ 0.001 Pa.s) \cite{Najafi2023Size-Surface}. The \texttt{time\_step} is set to 0.0005 s. \texttt{max\_displacement} of the AFM is 0.002 $\mu$m per \texttt{time\_step}, which means a maximum displacement of 4 $\mu$m per second, a velocity that had never been reached in our study. The bending \texttt{rigidity} of actin  $\kappa$ is set to 0.04 pN.$\mu$m, which means a persistence length $\ell_p \simeq$ 9.52 $\mu$m, in accordance with literature \cite{Gittes1993}. The segmentation $\mathrm{d}s$ is 0.03 $\mu$m, which means 10 points by filaments when they reach their maximum length. Nucleator stiffness of the link was arbitrarily set to 0.5 $\text{pN}.\mu \text{m}.rad^{-1}$ \cite{mund2018systematic}. 

\subsection{Activation of nucleators}
In simulations, the nucleator is a complex made of two domains. One binds existing actin filaments, the other nucleates a new filament. When a nucleator is added to the simulation, it starts with both domains unbound: this is the Free-Free (FF) configuration. Upon filament binding, it becomes Activated-Free (AF). It quickly nucleates a new filament to become Activated-Activated (AA). Binding follows effectively second-order reaction kinetics: each monomer can bind a filament with a rate $k_{on}=1000$ $\text{s}^{-1}$ when within a range $r_{on}=10$ nm of the filament. Nucleation follows first-order kinetics, as a bound nucleator will generate a new filament with a rate $k_{nuc} = 1000$ $\text{s}^{-1}$.
\medbreak 

\begin{figure}[h]
    \centering
    \includegraphics[width=0.9\linewidth]{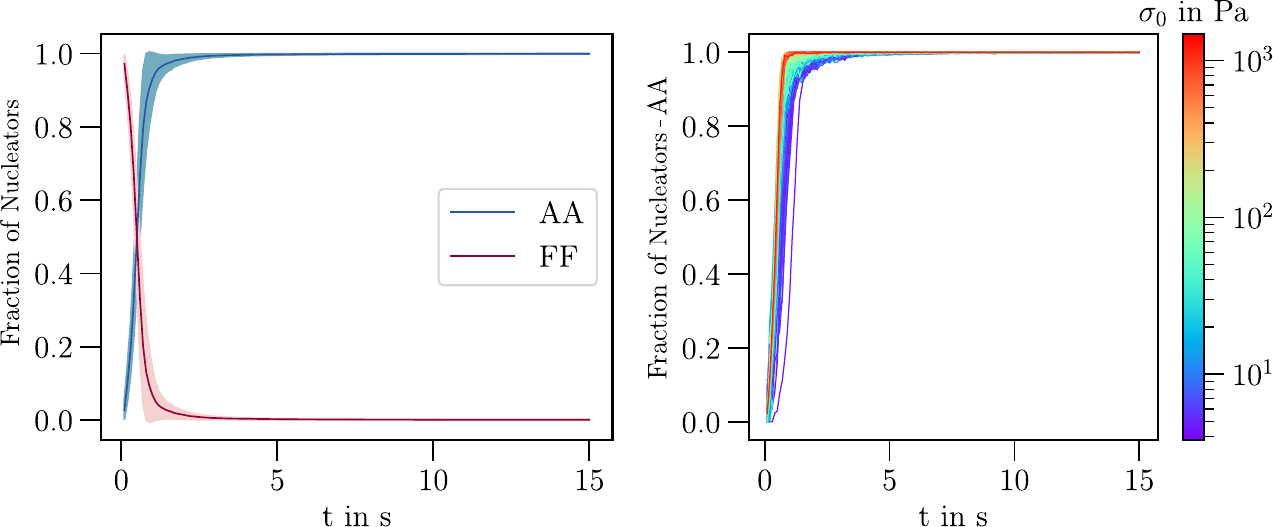}
    \caption{\centering{(Left) Mean fraction of nucleators AA (Activated-Activated) and FF (Free-Free) with respect to time, shaded part are standard deviation bounds. (Right) Detail on the AA curve for different stress values.}}
    \label{fig:s_5}
\end{figure}

Because $k_{on}$ is high, binding is fast as long as enough actin filaments are available. and the proportion of FF complexes therefore goes to 0 in about one second, FIG. \ref{fig:s_5} (Left), despite the regular addition of new FF complexes with rate $k$. Simultaneously, the proportion of AA complexes goes to 1. Because nucleation is fast, the proportion of AF complex is near-zero. Moreover, the time required to reach the plateau decreases with  $\sigma_0$ (FIG. \ref{fig:s_5}, Right), as higher stresses means a higher actin density, and thus faster nucleator binding.

\FloatBarrier 

\section{Stress adaptation}

\subsection{Estimation of velocity and density}

\begin{figure}[h]
    \centering
    \includegraphics[width=\linewidth]{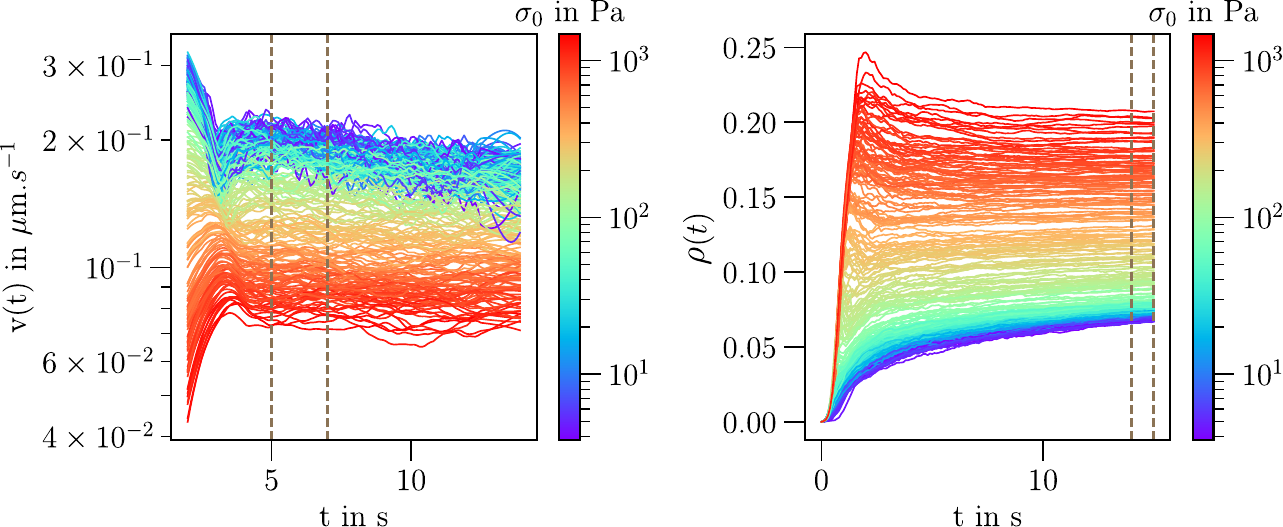}
    \caption{\centering{(Left) Instantaneous velocity fitted with a Savitzky-Golay filter as a function of time for different stress $\sigma_0$, with $f_s = \infty$ pN. Brown dotted lines delimit the fitting region, between 5 s and 7 s. (Right) Brown dotted lines delimit the fitting region for the estimation of the stationary density, between 14 s and 15 s.}}
\label{fig:s_6}
\end{figure}

In order to extract velocity from our position data we used a simple gradient method combined with a Savitzky-Golay filter. Basically, we define $n$ fitting intervals each centered on a data point $(x_0,y_0)$ (whenever it is possible) whose width is 2$\ell$ + 1. We also consider a polynomial of degree $d$, $P_d$ with $d < 2\ell + 1$. For each interval we do a regression and minimize the mean squared error to obtain the best fitting polynomial function $P_{d}^*$. Then we defined the smoothed value $y^*$ = $P_{d}^{*}(x_0)$. This technique has the advantage of being less abrupt than a simple rolling average. In FIG. \ref{fig:s_6} (Left) we plot the smoothed velocity for various $\sigma_0$ with parameters $p$ = 3 and $l$ = 20. Brown dotted lines delimit the fitting region to estimate the stationary velocity. For small $\sigma_0$ we can observe the slow down of the system which forces us to average for small time interval to neglect drag-induced effects. To estimate the stationary density we simply compute the average over the last frames, see FIG. \ref{fig:s_6} (Right).

\FloatBarrier

\begin{figure}[h]
    \centering
    \includegraphics[width=0.9\linewidth]{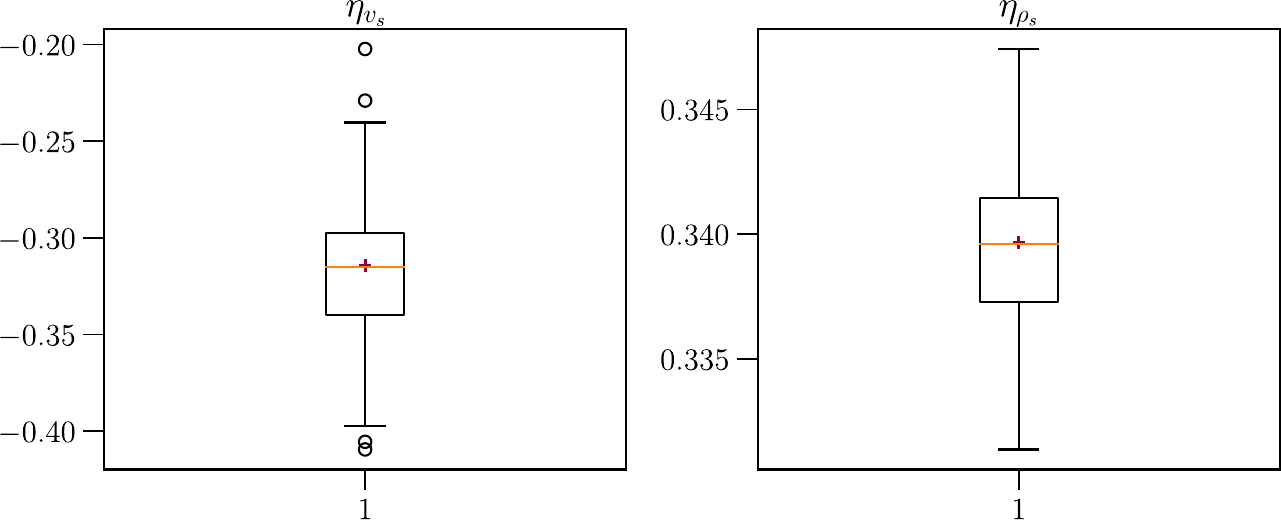}
    \caption{\centering{Bootstrap test on the exponent for velocity $\eta_{v_s}$ on (Left) and $\eta_{\rho_s}$ on (Right). The test has been done for $N_{sample}$ = 300 times. For each figure, the red cross is the original value while the orange line represents the bootstrap mean.}}
    \label{fig:s_7}
\end{figure}
\FloatBarrier

To establish the consistency of our measurements we performed a Bootstrap test on the velocity slope $\eta_{v_s}$ and on the density slope $\eta_{\rho_s}$. We sampled without replacement on the original fitting ensemble and obtained $N_{samples}$ = 300 bootstrap ensembles and conducted the same power law fitting for each. We represent the obtained box plots on FIG. \ref{fig:s_7}, for our infinite stall force simulations. Our original measurement (red cross) overlaps with the bootstrap mean (orange line).

\subsection{Flux of actin}

\begin{figure}[h]
    \centering
    \includegraphics[width=\linewidth]{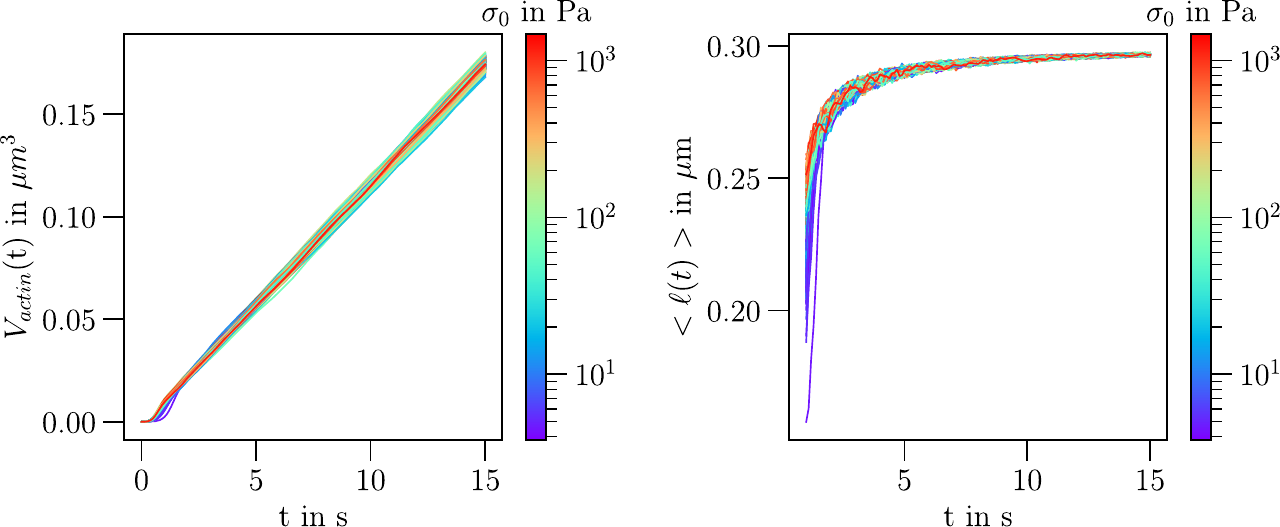}
    \caption{\centering{Total volume of actin $V_{actin}(t)$ on (Left) and mean length of filaments $ \langle \ell(t) \rangle$ (Right) as a function of time for different stress $\sigma_0$, with $f_s = \infty$ pN.}}
    \label{fig:s_8}
\end{figure}

From FIG. \ref{fig:s_8} (Left), we see that the volume of actin does not depend on the applied stress for infinite stall force filaments. From the (Right) one we conclude that the mean length of filaments $\langle\ell(t) \rangle$ reaches a value near its maximum possible one independently of the stress $\sigma_0$. 

\FloatBarrier

\section{Low stress regime}

\subsection{Drag estimation}

\begin{figure}[h]
    \centering
    \includegraphics[width=0.9\linewidth]{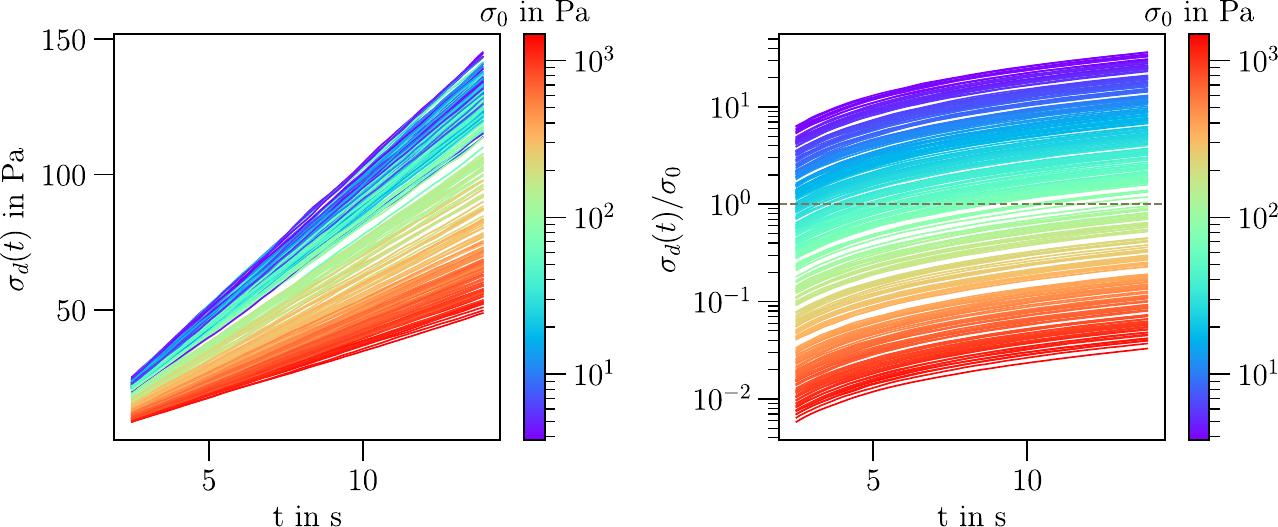}
    \caption{\centering{(Left) Drag stress $\sigma_{d}(t)$ as function of time $t$ for various stress $\sigma_0$. (Right) Ratio of the drag stress over the imposed stress: $\sigma_{d}(t)/\sigma_0$ as a function of time for various $\sigma_0$, with $f_s = \infty$ pN. The brown dotted-line represents the equality threshold.}}
    \label{fig:s_9}
\end{figure}

\subsection{Free growth and densification}

\begin{figure}[h]
    \centering
    \includegraphics[width=0.9\linewidth]{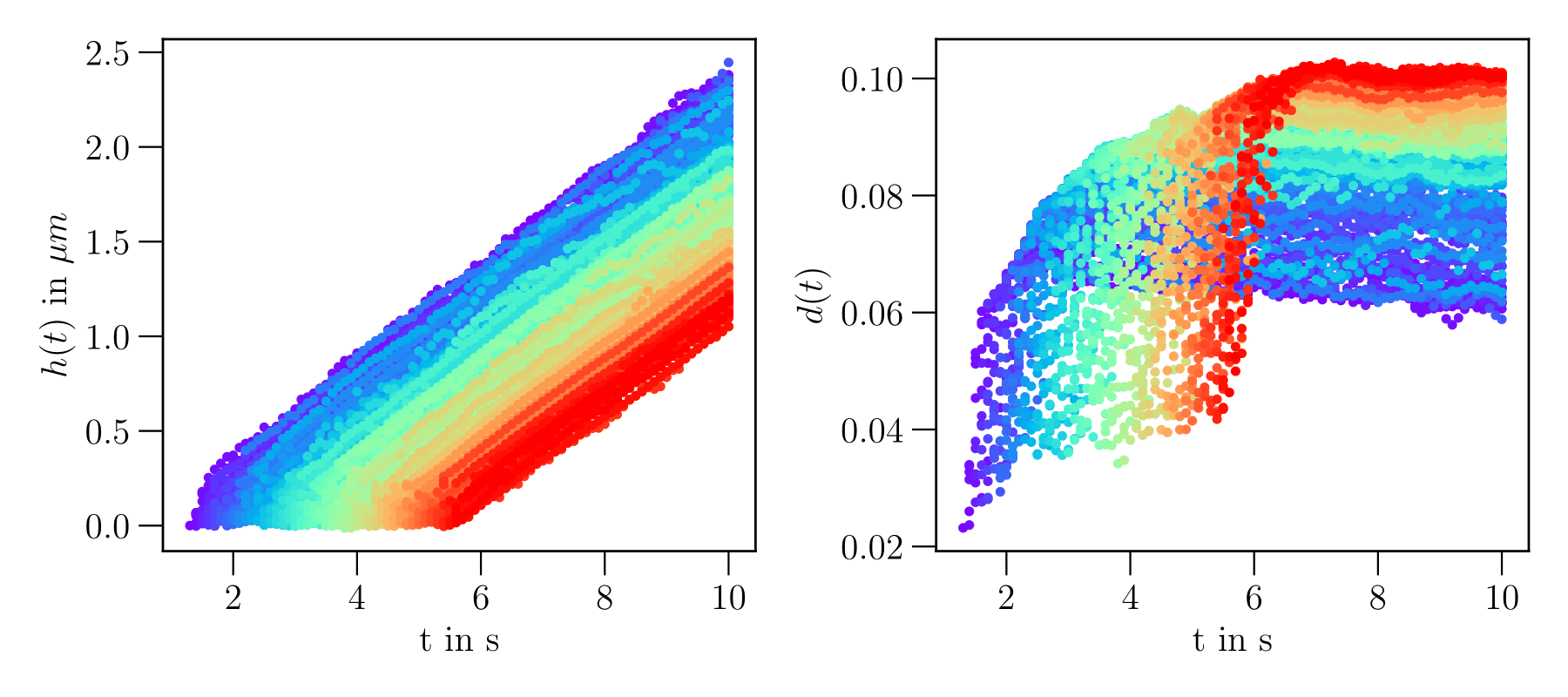}
    \caption{\centering{Evolution of slices with Height $h(t)$ (Left) and local density $d(t)$ on (Right) as a function of time $t$. The color code represents the appearance time. For each point the density is calculated inside the slice centered on the point position and with a height $\delta z$ = 0.2 $\mu$m. Here, $f_s = \infty$ pN and $k$ = 1378 $\mu$m\textsuperscript{-2}.s\textsuperscript{-1}}}
    \label{fig:s_10}
\end{figure}

\begin{figure}[h]
    \centering
    \includegraphics[width=0.9\linewidth]{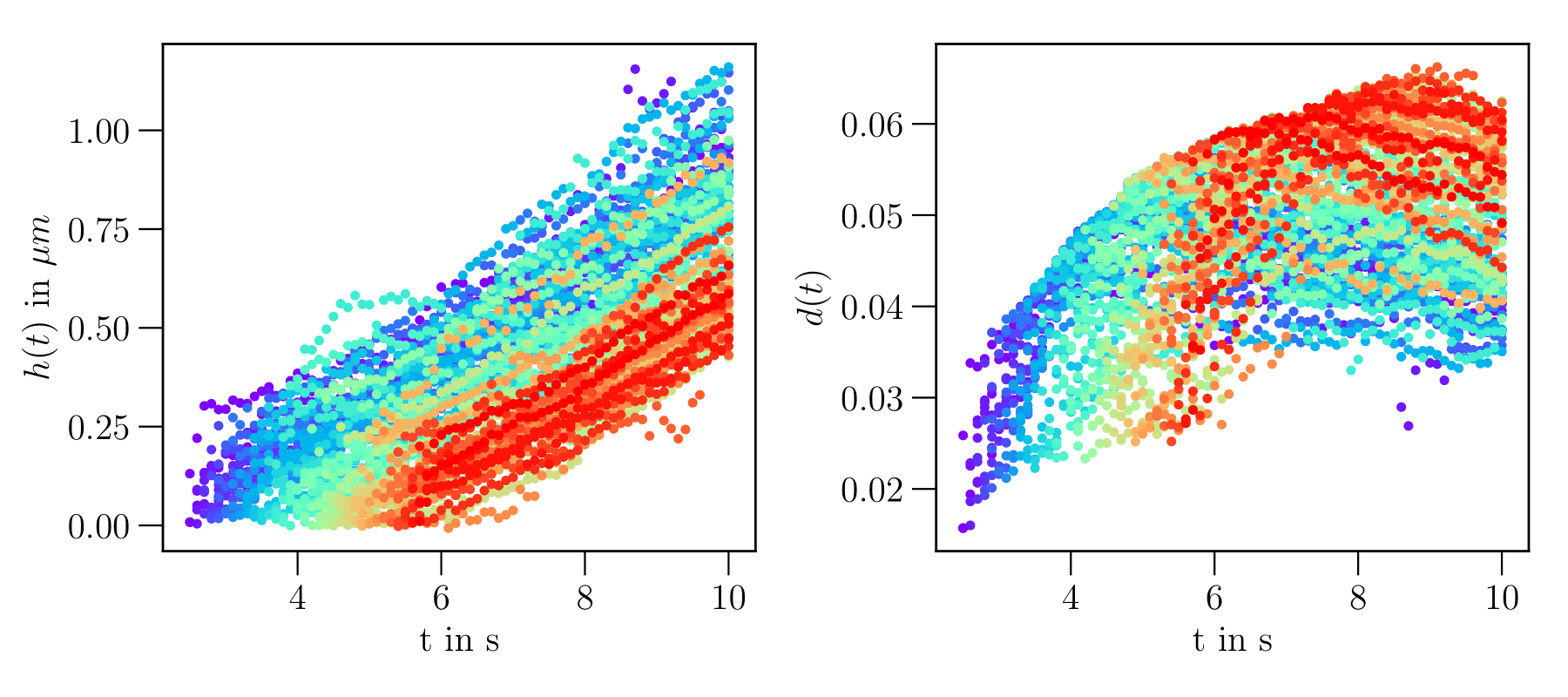}
    \caption{\centering{Evolution of slices with height $h(t)$ (Left) and local density $d(t)$ on (Right) as a function of time $t$. The color code represents the appearance time. For each point the density is calculated inside the slice centered on the point position and with a height $\delta z$ = 0.2 $\mu$m. Here, $f_s = \infty$ pN and $k$ = 390 $\mu$m\textsuperscript{-2}.s\textsuperscript{-1}}} \label{fig:s_11}
\end{figure}

The layering densification and the subsequent elastic relaxation are depicted in FIG \ref{fig:s_10} (Right) and \ref{fig:s_11} (Right). Here, each data point represents the average density within a volume centered on a segment of the filament. The color code denotes the time of appearance of each layer. On the left side of each figure, we depict the velocities of all filament segments, still labeled with their respective appearance times.

\FloatBarrier

\FloatBarrier

\subsection{Theoretical model}

The instantaneous velocity $v_{\tau}(t)$ of the network can be expressed as:

\begin{eqnarray}
    v_{\tau}(t) = v(t) + h(t) \bar{\dot{\epsilon}}(t) \ ; \ \bar{\dot{\epsilon}}(t) = \int_{0}^{h(t)}\frac{dx}{h(t)}\partial_{t}\epsilon(x,t)
\end{eqnarray}

Here, $v(t)$ is the polymerization-driven velocity, and $\bar{\dot{\epsilon}}$
is the average elastic relaxation rate across the network. As visible from FIG. \ref{fig:s_10} (Right), while density relaxation does occur, it can be reasonably neglected at the time scales of our observations, at least for high values of $k$. In fact, the most notable density relaxation observed is approximately 10\% over a time span of approximately 10 s. Thus, we assume that:

\begin{eqnarray}
    \vert h(t) \bar{\dot{\epsilon}}(t) \vert \ll v(t) \Rightarrow v_{\tau}(t) \simeq v(t)
\end{eqnarray}

In this approach, we conceptualize our system as comprising polymerizing filaments exerting force against a drag that progressively increases with time, denoted as $\sigma_{d}(t)$. Given that growing filaments ultimately reach their maximum length, the relationship between the drag $\sigma_{d}(t)$ and the velocity $v(t)$ becomes straightforward. Moreover, under the assumption that we are still operating within the entanglement regime, we can make use of entanglement equation. As a result, we employ the following equations:

\begin{gather} 
    \sigma_d(t) = \frac{\gamma_{d} N(t)}{S} v(t)\label{eq_sigma_d} \\
    v(t) = j \kappa^{2/7} r^{-8/7} \sigma_{d}(t)^{-2/7} \label{eq_martin_vd}  
\end{gather}

With  $j = k s_0 \ell_m$, and $N(t) = k S t$. $S$ is the surface of the enclosure and $s_0$ the cross section of a filament, $\gamma_0$ is the drag of a filament of length $\ell_m$:

\begin{eqnarray} \label{eq_gamma_0}
    \gamma_0 = \frac{3 \pi \eta \ell_{m}}{\ln \big(\frac{\ell_{m}}{2r}\big) + 0.312}
\end{eqnarray}

Then, mixing equations (\ref{eq_sigma_d}), (\ref{eq_martin_vd}) and (\ref{eq_gamma_0}), we derive expressions for growth velocity and density:

\begin{eqnarray}
    v(t) &=& t^{-2/9} \gamma_{0}^{-2/9} \kappa^{2/9} r^{8/9} k^{5/9} (s_{0}\ell_{m})^{7/9}  \\
    \rho(t) &=& t^{2/9} \gamma_{0}^{2/9} \kappa^{-2/9} r^{-8/9} k^{4/9} (s_{0}\ell_{m})^{2/9}
\end{eqnarray}

Thus, from previous equations we can derive a simple scaling of $v(t)$ and $\rho(t)$ with $k$ as the main parameter.

\begin{eqnarray} 
    v(t) &\propto& v_{d}(k) t^{-2/9} \\
    \rho(t) &\propto& \rho_{d}(k) t^{2/9}
\end{eqnarray}

This implies:

\begin{eqnarray}
    h(t) \propto \frac{9}{7} v_{d}(k) t^{7/9}
\end{eqnarray}

\subsection{Density scaling}

\begin{figure}[h]
    \centering
    \includegraphics[width=\linewidth]{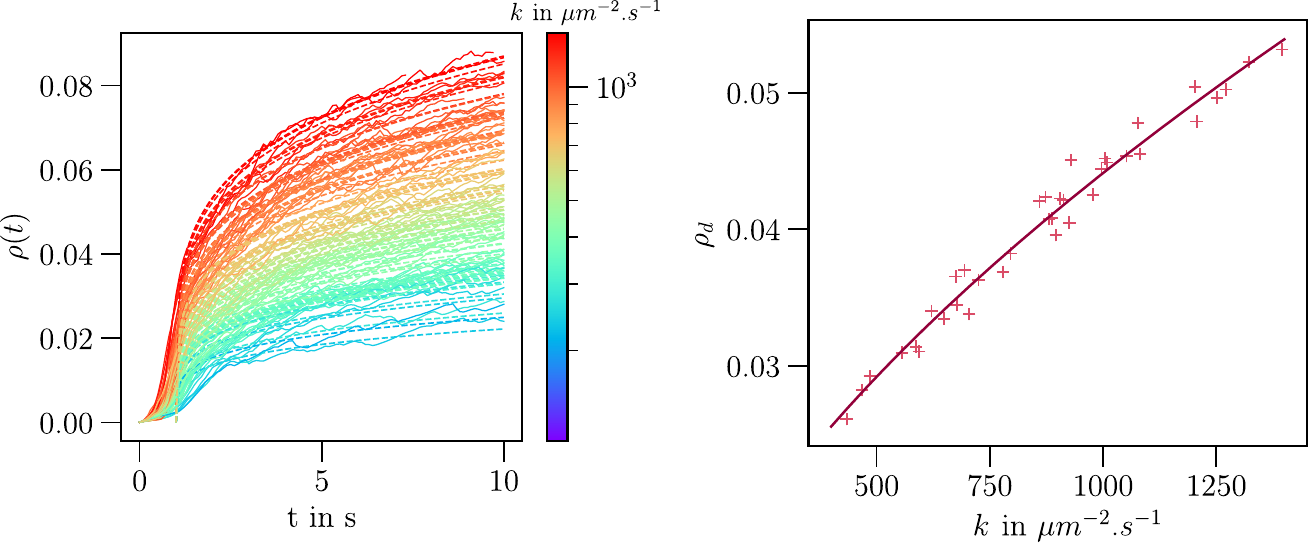}
    \caption{\centering{(Left) Density $\rho(t)$ as a function of time with the corresponding power-law fit (dotted line) for various nucleator rate $k$ and with $f_{s}$ = $\infty$ pN. (Right) Pre-factors of the power-law $\rho_d$ as a power law of $k$ with exponent 4/9 as expected from theory.}}
    \label{fig:s_12}
\end{figure}
\FloatBarrier
 
\section{Stall force dependence}

\subsection{Filament growth}
To simulate the Brownian ratchet dynamics, we incorporate the following expression for the growth of actin filaments within the Cytosim framework.

\begin{equation}  \label{numerical_br}
    v = v_0e^{-\frac{f}{f_0}} - v_{dep} \ \ ; \ \ f \ge 0
\end{equation}

$f_0$ is the \textit{growing force}, $v_0$ the unloaded polymerization velocity and $v_{dep}$ the depolymerization velocity. The stall force is thus:

\begin{eqnarray}\label{cyto_stall}
    f_{s} = f_0 \ln\Big(\frac{v_0}{v_{dep}}\Big)
\end{eqnarray}

Note that we also know that:  
\begin{eqnarray} \label{stall_b}
    f_{s} = \frac{k_BT}{a}\ln\Big(\frac{c}{c_{e}^0}\Big)
\end{eqnarray}
With $c_{e}^{0}$ the equilibrium concentration, with no force, and $c$ the actin concentration  \citep{Hill1981}.  We thus find \cite{badaoui2023multi}:
\begin{eqnarray} \label{physio_relevant_f_0}
    f_0 = \frac{k_{B}T}{a}
\end{eqnarray}

Thus, we can now compute the stall force corresponding to a physiologically relevant growing force $f_0$, as described in Equation \ref{physio_relevant_f_0}. Given $k_{B}T$ = 0.042 pN.$\mu$m and $a$ = 0.0025 $\mu$m (half the size of a G-Actin monomer), we determined $f_{s} \approx$ 3.87 pN.

\subsection{Simulation results}

\begin{figure}[h]
    \centering
    \includegraphics[width=\linewidth]{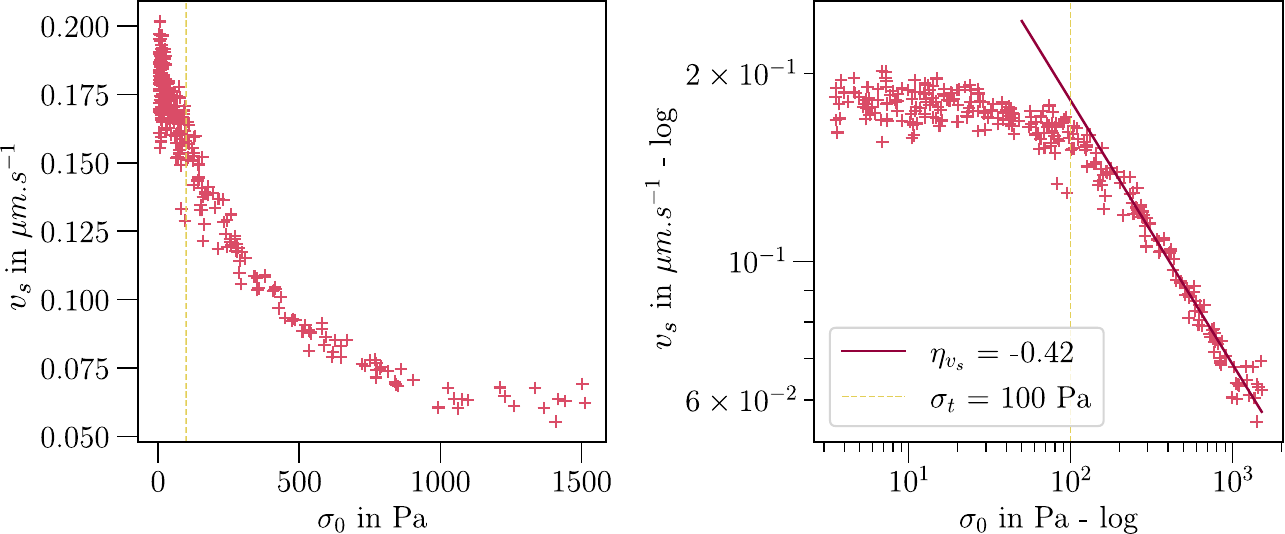}
    \caption{\centering{Stationary velocity $v_s$ for various stress $\sigma_0$ in normal (Left) and log-log scales (Right), with $f_{s}$ = 3.87 pN. The fit for the power law scaling is represented in the log-log scale.}}
    \label{fig:s_13}
\end{figure}